# Lithographyically defined, room temperature low threshold subwavelength red-emitting hybrid plasmonic lasers

Ning Liu[*†], Agnieszka Gocalinska[‡], John Justice[‡], Farzan Gity[‡], Ian Povey[‡], Brendan McCarthy[‡], Martyn Pemble[‡], Emanuele Pelucchi[‡], Hong Wei[§], Christophe Silien[†], Hongxing Xu[ǁ], Brian Corbett[*‡]

[†]Department of Physics and Bernal Institute, University of Limerick, Ireland
[‡]Tyndall National Institute, University College Cork, Ireland
[§]Beijing National Laboratory for Condensed Matter Physics, Institute of Physics, Chinese Academy of Sciences, Beijing 100190, China
[ǁ]School of Physics and Technology, and Institute for Advanced Studies and Center for Nanoscience and Nanotechnology, Wuhan University, Wuhan, 430072, China

Corresponding authors: ning.liu@ul.ie and brian.corbett@tyndall.ie

**Abstract:**

**Hybrid plasmonic lasers provide deep subwavelength optical confinement, strongly enhanced light-matter interaction and together with nanoscale footprint promise new applications in optical communication, bio-sensing and photolithography. The subwavelength hybrid plasmonic lasers reported so far often use bottom up grown nanowires, nanorods and nanosquares, making it difficult to integrate these devices into industry-relevant high density plasmonic circuits. Here, we report the first experimental demonstration of AlGaInP based, red-emitting hybrid plasmonic lasers at room temperature using lithography based fabrication processes. Resonant cavities with deep subwavelength 2D and 3D mode confinement of $\lambda^2/56$ and $\lambda^3/199$, respectively are demonstrated. A range of cavity geometries (waveguides, rings, squares and disks) show very low lasing thresholds of 0.6-1.8 mJ/cm² with wide gain bandwidth (610 nm-685 nm), which are attributed to the heterogeneous geometry of the gain material, the optimized etching technique, and the strong overlap of the gain material with the plasmonic modes. Most importantly, we establish the connection between mode confinements and enhanced absorption and**

**stimulated emission, which play a critical role in maintaining low lasing thresholds at extremely small hybrid plasmonic cavities. Our results pave the way for the further integration of dense arrays of hybrid plasmonic lasers with optical and electronic technology platforms.**

**TOC**

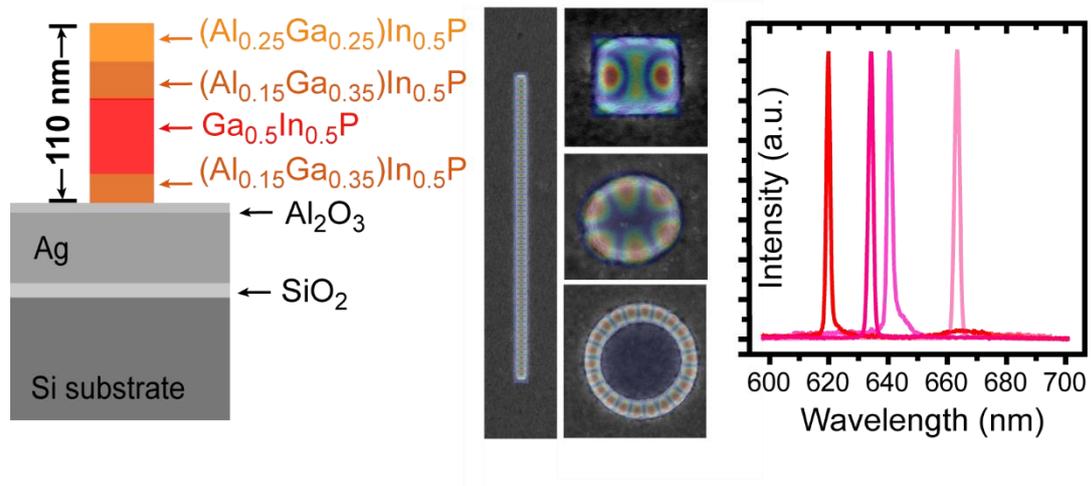

Integration has advanced the performance of photonics for applications in various areas of technology such as optical communications, sensing, biophotonics, and optical signal processing. [1-7] The footprint of an optical component is a critical parameter that determines the packing density of the devices and the overall performance of the chip. However, current optical components are incapable of achieving an integration density similar to that of integrated electronics due to the diffraction limit of light. Different designs of cavities, including metallo-dielectric cavities and photonic crystals have been explored to confine light into a subwavelength (vacuum) volume, however the volume is still limited by diffraction ~ ($\lambda/2n$)$^3$. [8-10] To overcome this limitation, a special optical mode, commonly known as the surface plasmon mode corresponding to coherent oscillations of electrons at the metal-dielectric interface, has been exploited.[11] As the electric field associated with surface plasmon modes exponentially decays into both metal and dielectric media, it has been widely demonstrated that surface plasmon modes can be confined into the deep subwavelength domain to the extent of $\lambda/20$ thus making manipulating light at the nanoscale both possible and beneficial.[12-18]

In particular, hybrid surface plasmon modes, which take advantage of an ultrathin low refractive index insulator layer between the metal and dielectric material, demonstrate both deep subwavelength mode confinement and relatively low propagation losses, and as a result have recently attracted great attention.[19,20] Since the first experimental demonstration of hybrid plasmonic lasers by Oulton *et al.*,[20] the emission wavelengths of plasmonic lasers have spanned from 370 nm to 650 nm, and the optically pumped lasing threshold of these devices at room temperature has decreased from ~2 GW/cm$^2$ down to ~3 MW/cm$^2$.[21-25] Nevertheless, the subwavelength hybrid plasmonic lasers reported thus far use the natural growth geometries of nanowires, nanorods, and nanosquares.[20-23,25] Furthermore, control over the position and the size of the devices needs to be addressed before hybrid surface plasmonic lasers can be used in industry-relevant applications.

Here, we report the first experimental demonstration using lithography based fabrication approach of red-emitting (610nm to 685 nm) deep sub-wavelength room temperature hybrid plasmonic laser arrays. We demonstrate lasers with different cavity geometries, including hybrid plasmonic waveguides and square cavities, which share similarities to those demonstrated using bottom-up semiconductor nanostructures, as well as hybrid plasmonic disk and ring lasers, which have not been previously demonstrated at such length scales. These hybrid plasmonic lasers exhibit very low lasing thresholds at room temperature, namely 0.6 mJ/cm$^2$ to 1.8 mJ/cm$^2$ per pulse when pumped by a laser with pulse duration of 4 ps, or with a peak pump intensity down to 14 MW/cm$^2$ when pumped by a laser with pulse duration of 76 ps, along with wide lasing bandwidth (610 nm to 685 nm). The reason for the outstanding performance of these devices is multifold, including the heterogeneous geometry and optimized etching technique for high optical gain and long radiative lifetime, the improved alignment between the mode distributions and gain material, the decreased ohmic loss due to the higher refractive index of the semiconductor material, and the enhanced absorption and stimulated emission due to the excitation of hybrid plasmonic modes.

**Results**

Recently, GaN based semiconductors have demonstrated both high optical gain and relatively low lasing thresholds in UV and visible emitting hybrid plasmonic lasers using as-grown nanowires and nanorods.[22,23,25] In our lithography based approach we extend the gain material to the $Al_xGa_{0.5-x}In_{0.5}P$ system where an engineered epitaxial heterostructure with a total thickness of 110 nm is used. This material has demonstrated very high optical gain and wide gain bandwidth at room temperature.[26] The layered gain material was grown on an $Al_{0.75}Ga_{0.25}As$ etch stop layer on a GaAs substrate by metal organic vapour phase epitaxy. To limit the non-radiative surface recombination the top and bottom layers of the heterostructure are wider bandgap $(Al_{0.3}Ga_{0.7})_{0.5}In_{0.5}P$ material to confine carriers within a 50 nm thick $Ga_{0.5}In_{0.5}P$ gain layer. As a result, only the edges (perimeters) of the nanostructures are susceptible to surface recombination. After the nanostructures were defined by electron beam lithography, the devices were etched by $SiCl_4$/He inductively coupled plasma etching. Photoluminescence (PL) lifetime measurements of the semiconductor structures by $SiCl_4$-based plasma etching show a total lifetime of 360 picoseconds, a significant increase in performance compared to the 53 picoseconds fast recombination time by HBr-based wet etching, attributed to non-radiative surface recombination[27] (see SI-2 for details). The semiconductor nanostructures were then covered in black wax and transferred to $Ag/SiO_2$/Si substrate, separated by an ultrathin layer of $Al_2O_3$ of 6 nm (see SI Methods for details). The final layout of the nanostructures is described in Figure 1a, with SEM images of representative nanostructure arrays of different cavity geometries given in Figure 1b-1e. The semiconductor nanostructures were also transferred to a glass coverslip for comparison of the device performance.

**Hybrid plasmonic waveguide and ring lasers with 2D mode confinement**

All optical measurements were carried out at room temperature in ambient condition and on individual devices. As an example of hybrid plasmonic lasers with 2D confinement, we fabricated waveguides with various widths and lengths. These hybrid plasmonic lasers show very low lasing threshold. We observe lasing from all waveguides, including the smallest 150 nm × 8 μm (width × length) waveguides, whose lasing occurs around 1.3±0.2 mJ/cm$^2$, as shown in Figure 2a, when pumped by a picosecond laser at 570 nm (4 ps, 1 kHz). Polarization studies on the emitted light confirms that the lasing output is mostly along the long axis of the waveguide, suggesting lasing due to the hybrid plasmonic modes (see SI-3 for details). As the pump intensity increases, multiple Fabry-Perot cavity modes are observed in the spectra (see SI-4 for details). From the spacing between adjacent cavity modes, we can estimate the effective group refractive index $n_g$ of the optical modes. Fitting the data to $\Delta\lambda = \lambda^2/2n_g l$, where $l$ is the length of the waveguide, yields $n_g$=6.3, a value much higher than $n_g$=4.0 from reference [28], revealing strong dispersive effects on the refractive index due to band filling and free carrier effects at the high photo-generated carrier density (see SI-5 for details). The low lasing threshold observed can be partially attributed to the improved mode overlap with the gain material and the close contact of the semiconductor to the insulator/metal substrate due to the top-down fabrication applied. The planar contact geometry of the nanostructures with the substrate maximizes the hybrid plasmonic mode overlap with the semiconductor gain material when compared with that of nanowires with circular cross section. The electric field distribution at the plane perpendicular to the long axis of the hybrid plasmonic waveguide was simulated using the COMSOL Multiphysics package and is shown in Figure 2b. If we define the mode overlap factor of the gain material $\Gamma_s$ as the ratio of the electric energy in the gain region over the total energy of the mode $\Gamma_s = \frac{\iiint_{semicon} W(r)d^3r}{\iiint_{total} W(r)d^3r}$, where $W(\mathbf{r})$ is the electric energy density of the mode, the mode overlap factor in current configuration yields a value of 0.89, which is nearly twice of that of a circular cross section (0.4 to 0.52 in ref 20). In addition, it has been shown that a planar contact geometry can help to improve the exciton–surface plasmon energy transfer.[23] The high refractive index of $Al_xGa_{0.5-x}In_{0.5}P$ (n=3.4 to 3.6, depending on the composition and wavelength) also effectively reduces the amount of electric energy within the metal substrate (the overlap factor for metal $\Gamma_m$ = 2.3 × 10$^{-4}$ in the current case), so that the overall propagation loss of the hybrid plasmonic mode is substantially decreased to 3273 cm$^{-1}$ (see SI-6 for details), only 53% of the reported propagation loss (6230 cm$^{-1}$) of a hybrid plasmonic waveguide formed by a CdSe nanobelt (n=2.7) of a similar geometry on an $Al_2O_3$/Ag substrate.[29]

To demonstrate the versatility of our lithography based approach, we fabricated ring cavities of various sizes. This geometry is used in conventional micro photonics because of its low loss and ease of coupling to waveguided light.[30,31] However, the ring geometry could not be demonstrated in hybrid plasmonic lasers since as-grown nanomaterials can hardly form this shape. Figure 2c shows the PL evolution of one of the smallest hybrid plasmonic ring lasers we fabricated (outer diameter: 1.09 μm, inner diameter: 0.79 μm) with increase of pump intensity. Lasing from two whispering gallery modes (610 nm and 634 nm) is observed at a pump energy of 2.2 mJ/cm$^2$, slightly higher than that of the hybrid plasmonic waveguide laser (Figure 2a-b) of similar geometry in cross section. Even though the whispering gallery modes satisfy total reflection condition, the increase in radiative loss as the feature size decreases leads to higher overall loss of the hybrid plasmonic ring modes. Figure 2e-2g show the simulated electric field |E| distribution of the whispering gallery modes $W_{8,2}$, $W_{11,1}$, and $W_{7,2}$ at 610 nm, 634 nm and 635 nm, respectively. $W_{11,1}$, and $W_{7,2}$ are close in both wavelength and mode loss and we cannot tell them apart with our current set-up. The simulated mode frequencies match the experimental results the best for carrier density $N$ = 1.0 × 10$^{19}$ cm$^{-3}$, which causes a significant change in the absorption coefficient $\alpha$, leading to a blue shift of resonance modes by 5 nm (see SI-5 for details). The hybrid plasmonic waveguide and ring lasers demonstrate very strong 2D confinement. The effective area of the hybrid plasmonic mode $\bar{A} = \frac{\iint W(r)d^2r}{Max(W(r))}$ is around $\lambda^2/56$, or 0.32($\lambda/2n_{\text{eff}}$)$^2$ for the waveguide shown in Figure 2a.

**Hybrid plasmonic circular disk and square lasers with 3D mode confinement**

The packing density of optical devices, however, is ultimately determined by the 3D confinement of the optical mode. We investigated squares and disk cavities which are chosen to satisfy total internal reflection condition in order to achieve low overall loss. Figure 3a-3b show the PL evolution of two hybrid plasmonic disk cavities and Figure 3e-3f show that of two slightly elongated hybrid plasmonic square cavities respectively. It is clearly shown in the figure that lasing behaviour is observed accompanied by the narrowing of the FWHM of the resonant peak as the pump energy increases. Different from the multimode lasing behaviours of hybrid plasmonic waveguide and ring cavities, we observe only single mode lasing as the lateral dimension of the disk and rectangular cavities decreases below 570 nm. In addition, we also notice that the 'kink' associated with the transition from spontaneous emission to stimulated emission in the PL intensity *vs*. pump energy plot in Figure 3 becomes less distinguishable at the smaller cavity size of 350 nm × 390 nm (short axis × long axis). We simulated the exciton density and optical mode intensity within the cavity using a pair of coupled rate equations (see SI-9 for details). The spontaneous emission coefficient $\beta$ is used to indicate the effectiveness of the spontaneous emission coupled to a specific cavity mode. Fitting the output intensity-pump energy data yields $\beta$ values of 0.02±0.01, 0.01±0.005, 0.25±0.07, and 0.5±0.1 for Figure 3a-3b and Figure 3e-3f, respectively. Both analytical solution[21,32,33] and the COMSOL simulation demonstrate that as the dimensions of the cavities decrease, the free spectral range between two adjacent resonant cavity mode increases. For a rectangle cavity of 350 nm × 390 nm, there are only two cavity modes within the spectral range of 610 nm to 685 nm. It is clearly shown in Figure 3f that the spontaneous emission effectively couples into a cavity mode at 650 nm and the emission peak narrows down uniformly with the pump intensity. This is different from the PL evolution shown in Figure 3e, where the emission band around 670 nm includes multiple cavity modes. As the pump intensity increases, a specific mode gets effectively amplified while the spontaneous emission from other modes within the band can still be observed. Similar results are also obtained for the disk cavity with a dimension of 350 nm × 390 nm (short axis × long axis). The effective mode volume of the hybrid plasmonic rectangle cavity of 350 nm × 390 nm at 640 nm is around $\lambda^3/199$, i.e. $0.21(\lambda/2n_{\text{eff}})^3$.

**Comparison of the Purcell effect on 3D hybrid plasmonic and photonic nanostructures**

It has been widely discussed in literature that the confinement in hybrid plasmonic cavities can substantially enhance the spontaneous emission rate due to the Purcell effect. The enhancement factor varies from 2 to 20, depending on the specific geometry of the cavity. Unlike nanowire cases, whose mode confinement is achieved in two dimensions, the Purcell factor of hybrid plasmonic cavities with mode confinement in all three dimensions is less studied due to the limited control over the size of as-grown nanostructures. With our top-down approach, we could study the change of PL lifetime and therefore the Purcell factor as the size of cavity shrinks. The lifetimes of all nanostructures are measured using time correlated single photon counting method. Figure 4a shows the PL lifetimes of a hybrid plasmonic disk of 350 nm × 390 nm × 110 nm (long axis × short axis × height) and that of its photonic counterpart of the same size, which are 14 ± 4 ps and 158 ± 4 ps accordingly, after deconvolution from the instrumental profile (see SI-2 for details). Compared to the spontaneous emission lifetime of 350 ± 8 ps obtained on the gain material of large size and released on glass, the Purcell factor due to the strong mode confinement is estimated to be 25 for the hybrid plasmonic cavity and 2.2 for the photonic cavity. Figure 4b gives the summary of Purcell factors of all 3D cavities obtained from their measured PL lifetimes. It is clear from Figure 4b that the 3D hybrid plasmonic cavities exhibit strong mode confinement as the size of the cavity decreases, which leads to a Purcell factor over 29 at the hybrid plasmonic rectangle cavity of 350 nm × 390 nm × 110 nm, one of the highest values reported so far. 3D photonic cavities, on the other hand, only show moderate enhancement, with a maximum Purcell factor of 4.3 at the rectangle cavity of 350 nm × 390 nm × 110 nm. These values not only influence the spontaneous emission rate, but also have direct impact on the lasing threshold of the cavities.

**Discussion**

The lasing thresholds of hybrid plasmonic cavities are compared in Figure 5a with those of photonic cavities. For disk shaped cavities, the lasing threshold for photonic cavities is lower than that of the hybrid plasmonic cavities at relatively large diameters (> 1 μm). However, no lasing is observed for photonic cavities of diameter smaller than 570 nm, while lasing is still achieved for hybrid plasmonic cavities of diameter ranging from 370 to 570 nm. The simulated cavity loss is plotted as a function of the disk diameter for both photonic and plasmonic cases of circular shape in Figure 5b, where the mode of least loss for a given diameter is chosen in the wavelength range of 610 nm to 685 nm. We can see that the losses of the photonic modes are much lower than those of the hybrid plasmonic cavities at relatively large diameters. However, the loss increases rapidly as the size of the cavity decreases below the wavelength of the resonant cavity mode due to the loss of mode confinement and increased radiative loss. The crossover happens at the diameter around 650 nm, below which lasing is hard to be sustained in the pure photonic cavity.

It should be noted that the lasing threshold of hybrid plasmonic lasers at the smaller dimensions of 370 to 570 nm is lower than that of the larger ones (Figure 5a). This cannot be explained by the largely monotonically increase in cavity losses with decreasing size. We attribute the opposite trend in lasing threshold vs. the cavity loss to the enhanced absorption and enhanced stimulated emission due to the Purcell effect (see SI-7-11 for details). Figure 5c and 5d show the simulated electric field intensity $|E|^2$ distribution within hybrid plasmonic disk cavities of a diameter of 350 nm and a diameter of 1100 nm at the excitation wavelength of 570 nm, respectively. It is clear from the simulation that hybrid plasmonic modes are excited in the case of smaller dimension, which leads to a strong enhancement in the electric field within the gain region. The averaged $|E|^2$ over the volume of semiconductors is 2.97 for Figure 5c, 6.5 times of that for Figure 5d (0.46) under the same excitation conditions. The higher electric field intensity within the semiconductor enhances the absorption of light, which results in a higher carrier density within the gain region and a lower lasing threshold.

The enhanced absorption alone cannot quantitatively match well with our experimental values. Enhanced stimulated emission also needs to be considered in this case. Though much has been discussed on the relationship between Purcell effect and spontaneous emission rate, less attention is paid on the influence of Purcell effect on the stimulated emission. Einstein's relationship dictates that the stimulated emission cross section $\sigma_{st}$ is linearly proportional to the spontaneous emission rate $\gamma_{sp}$. The stimulated emission cross section $\sigma_{st}$ should therefore be enhanced by the same Purcell factor as that for $\gamma_{sp}$ (see SI-9 for details). Since the gain coefficient $g$ is related to $\sigma_{st}$ through $g=\sigma_{st}(N-N_{th})$, where $N$ is the photo-generated carrier density and $N_{th}$ the required carrier density to reach transparency threshold, this means that an enhanced $\sigma_{st}$ needs smaller $N$ to reach lasing threshold. This effect was observed in photonic micro-disks and micro-droplets[34,35] and the possibility of observing similar phenomena in a plasmonic cavity was discussed by Lawrence *et al.*[36] To simulate the lasing thresholds for both photonic and hybrid plasmonic cases, we take into account both enhanced absorption and enhanced stimulated emission in the coupled rate equations (See SI-11 for details). The simulation results are also plotted in Figure 5a and yield reasonably good agreements with our experimental values. It is noted that as the emission band of the gain material is from 610 nm to 685 nm and lasing can only occur within this wavelength range, the cavity mode of the least loss can swing in and out of this range with the decreasing of cavity size. This leads to the discontinuity in cavity loss of the optimal mode in this wavelength range as a function of the cavity size, as shown in Figure 5b. As a result, the simulated lasing thresholds show oscillatory behaviour (Figure 5a).

The experimental results discussed so far are obtained using a 4 ps pulsed laser. To understand the role of pulse width in achieving the low threshold of the hybrid plasmonic lasers, we also measured the lasing behaviour of our nanostructures using a supercontinuum laser with a typical pulse width of 76 ps. A bandpass filter (central wavelength 575 nm, bandwidth 25 nm) was used to limit the pumping wavelength. The measured lasing thresholds were also shown in Figure 5a. We found that lasing is achieved at consistently higher pulse energy by 8% to 70% compared to that by the 4ps pulsed laser, but with lower peak pump intensity. The model based on rate equations is again used to evaluate the change of threshold with the increase of pump pulse width, as given in Figure 5e (see SI-11 for details). It is predicted from our simple model that the lasing threshold energy increases monotonically with the increase of pulse duration, while the pump intensity threshold exhibits a minimum at the optimized pump duration, in this case around 100 ps, directly related to the non-radiative lifetime.    .

In conclusion, we have demonstrated room temperature, low threshold red-wavelength lasing from hybrid plasmonic lasers of various cavity shapes and sizes using a lithography based fabrication technique. This approach allows for the realisation of high density plasmonic circuits on a single chip that can be easily integrated with electronics or other waveguide platforms. Our study also reveals the important relationship among Purcell effect, pulse duration and the lasing threshold, where further decrease in the intensity of lasing threshold of hybrid plasmonic lasers is possible if cavity modes can be further confined and if the non-radiative lifetime can be increased by effective surface passivation.

**Acknowledgments**

N.L. thanks the support by Science Foundation Ireland National Access Programme (No. 444). E.P. acknowledges the funding provided by Science Foundation Ireland under grants 12/RC/2276 and 10/IN.1/I3000. I.P. and M.P. acknowledge the funding provided by Science Foundation Ireland under grants 11/PI/1117. H.W. and H.X.X thank the support by Ministry of Science and Technology of China (Grant No. 2015CB932400), and National Natural Science Foundation of China (Grant Nos. 11134013 and 11227407).

# Figures

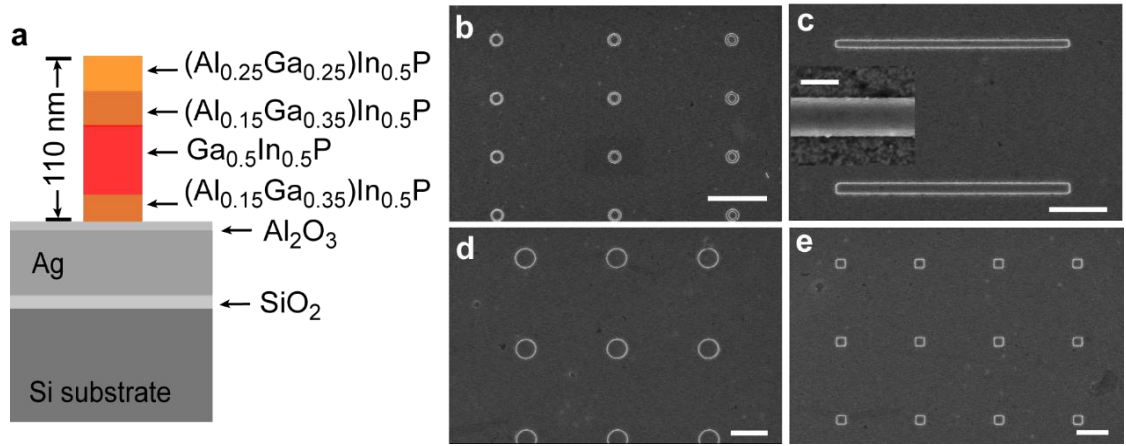

**Figure 1. Device configuration and SEM images of released devices (a).** Schematic of the cross section of hybrid plasmonic cavities fabricated using top-down lithography. The thicknesses of $(Al_{0.25}Ga_{0.25})In_{0.5}P$, $(Al_{0.15}Ga_{0.35})In_{0.5}P$, $Ga_{0.5}In_{0.5}P$, $Al_2O_3$, Ag, $SiO_2$ are 20 nm, 20 nm, 50 nm, 6 nm, 70 nm, and 25 nm, respectively. **(b-e).** Scanning electron micrographs of transferred rings, waveguides, circular disks, and square shapes. The scale bar is 5 μm in **(b)**, 2 μm in **(c-e)** and 150 nm in the inset of **(c)**.

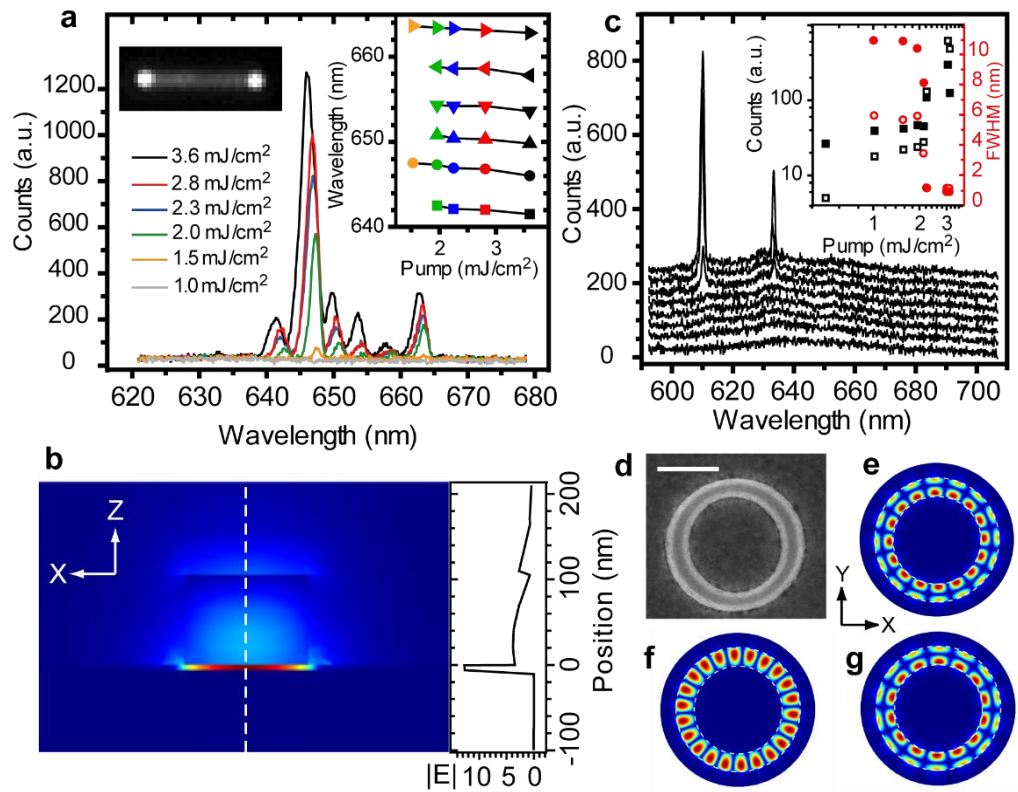

**Figure 2. Hybrid plasmonic lasers with 2D confinement. (a).** Lasing output from a 110 nm × 150 nm × 8 μm (height × width × length) waveguide with the increase of pump energy. Insets show an optical image of the waveguide recorded with a 600 nm long pass filter when pumped at 3.6 mJ/cm$^2$ and the blue shift of Fabry-Perot resonant cavity wavelengths with the increasing pump energy. **(b)**. COMSOL simulation of electric field amplitude distribution in the plane perpendicular to the long axis of the waveguide and a line cut showing the |E| along the dashed line. **(c)**. PL output of a hybrid ring cavity at different pump energy. The inset shows the output intensity at 610 nm (hollow black squares) and 635 nm (solid black squares), as well as the full width at half maximum (FWHM) of the peaks at 610 nm (hollow red circles) and 635 nm (solid red circles) with the increase of pump energy. The spectra are offset for clarity. **(d)**. Scanning electron micrograph of the hybrid ring cavity (outer diameter: 1.09 μm, inner diameter: 0.79 μm, width of the ring 150 nm), with scale bar corresponding to 500 nm. **(e-g)**. COMSOL simulations of electric field amplitude |E| distribution at the middle of the $Al_2O_3$ layer, for whispering gallery modes $W_{8,2}$, $W_{11,1}$, $W_{7,2}$ at 610 nm, 634 nm and 635 nm respectively. The dashed curves describe the contour of the ring.

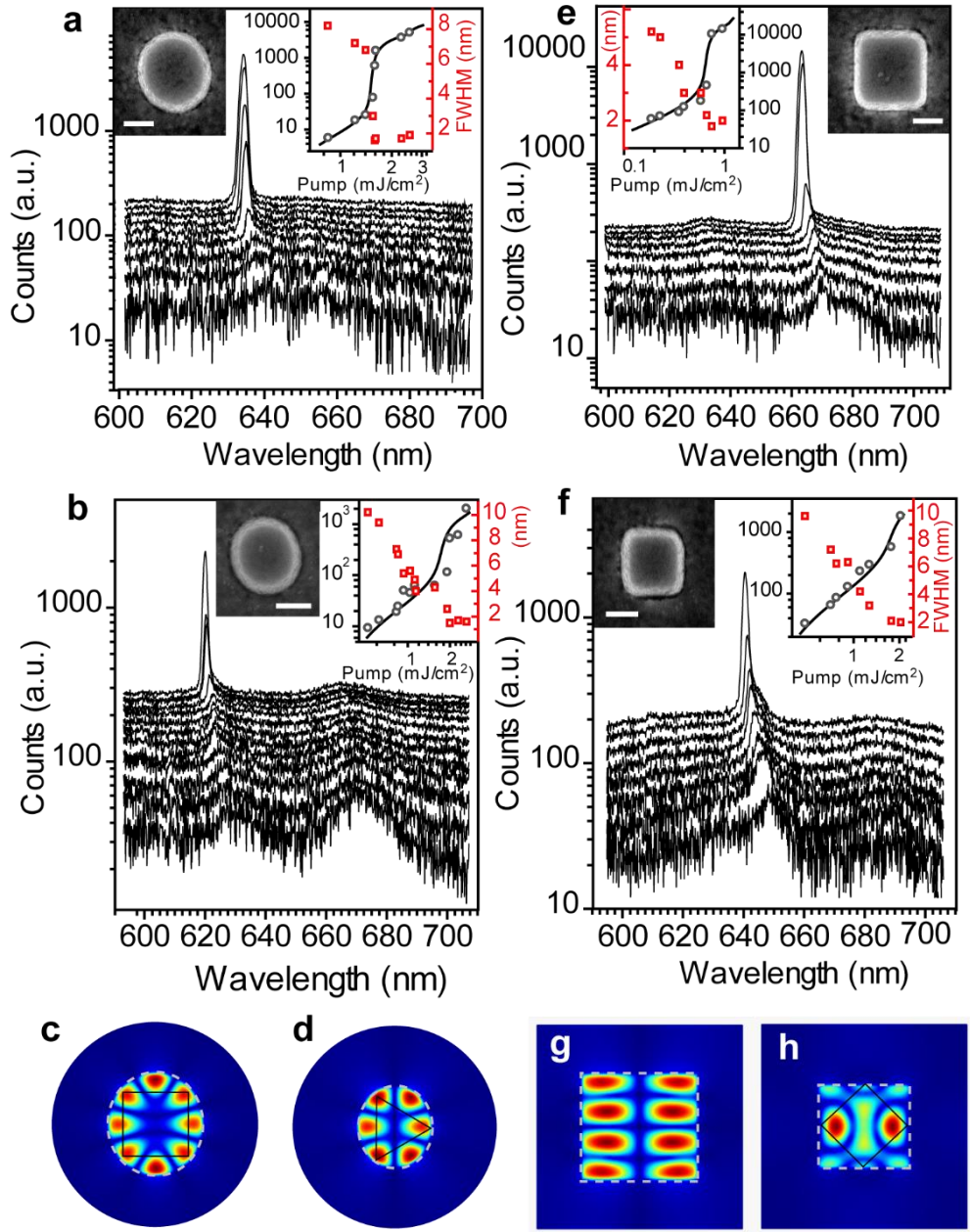

**Figure 3**. **Hybrid plasmonic lasers with 3D confinement. (a-b).** PL evolution of slightly elongated hybrid plasmonic circular disks of 450 nm × 490 nm and 350 nm × 390 nm as a function of pump energy, respectively. All spectra are offset for clarity. The insets show the SEM images of the nanolasers and the intensity output at the resonance peak with the change of pump energy (grey circles) and fitting to the rate equations detailed in SI-9. The FWHMs of the resonance peak (red squares) are also plotted as the function of the pump energy. The scale bars in the SEM images are 200 nm. **(c-d)**. COMSOL simulation results of electric field intensity $|E|^2$ distribution at the middle of $Al_2O_3$ layer of the resonant whispering gallery mode at 634 nm and 620 nm for **(a)** and **(b)**, respectively. The white dashed curves show the physical contours of the disks and black solid lines indicate the traveling path in ray optics. **(e-f).** PL evolution of rectangle hybrid plasmonic cavities of 450 nm × 490 nm and 350 nm × 390 nm as a function of pump energy, respectively. All spectra are offset for clarity. The scale bars are 200 nm and the insets show the intensity output at the resonant peak with the change of pump energy (grey circles) and fitting to the rate equations detailed in SI-9. The FWHMs of the resonance peak (red squares) are also plotted as the function of the pump energy. **(g-h)**. COMSOL simulation results of electric field intensity $|E|^2$ distribution at the middle of $Al_2O_3$ layer of the resonant whisper gallery mode at 663 nm and 640 nm for **(e)** and **(f)**, respectively. The white dashed curves show the physical contours of the rectangles and black solid lines indicate the travelling path in ray optics. Note that the lasing mode in 450 nm × 490 nm rectangle cavity is a Fabry-Perot type of mode with standing wave distribution along the edges of the cavity.

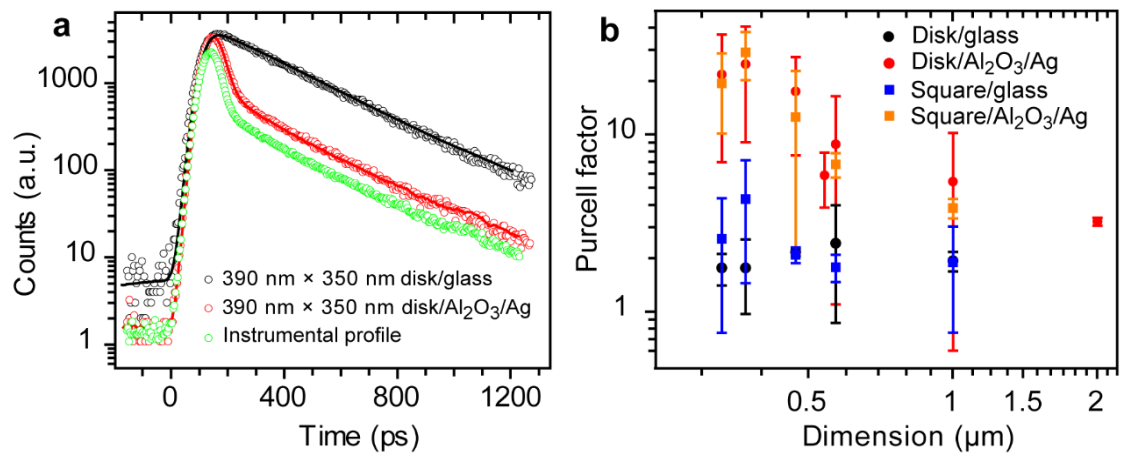

**Figure 4**. **Lifetime measurement and Purcell effect of hybrid plasmonic and photonic cavities with 3D mode confinement. (a)** Lifetime measurements on one of the smallest hybrid plasmonic disk cavities and photonic disk cavities. **(b)** Purcell factors of hybrid plasmonic and photonic cavities as a function of cavity dimension. (The size of the cavity is taken as the average of long and short axis)

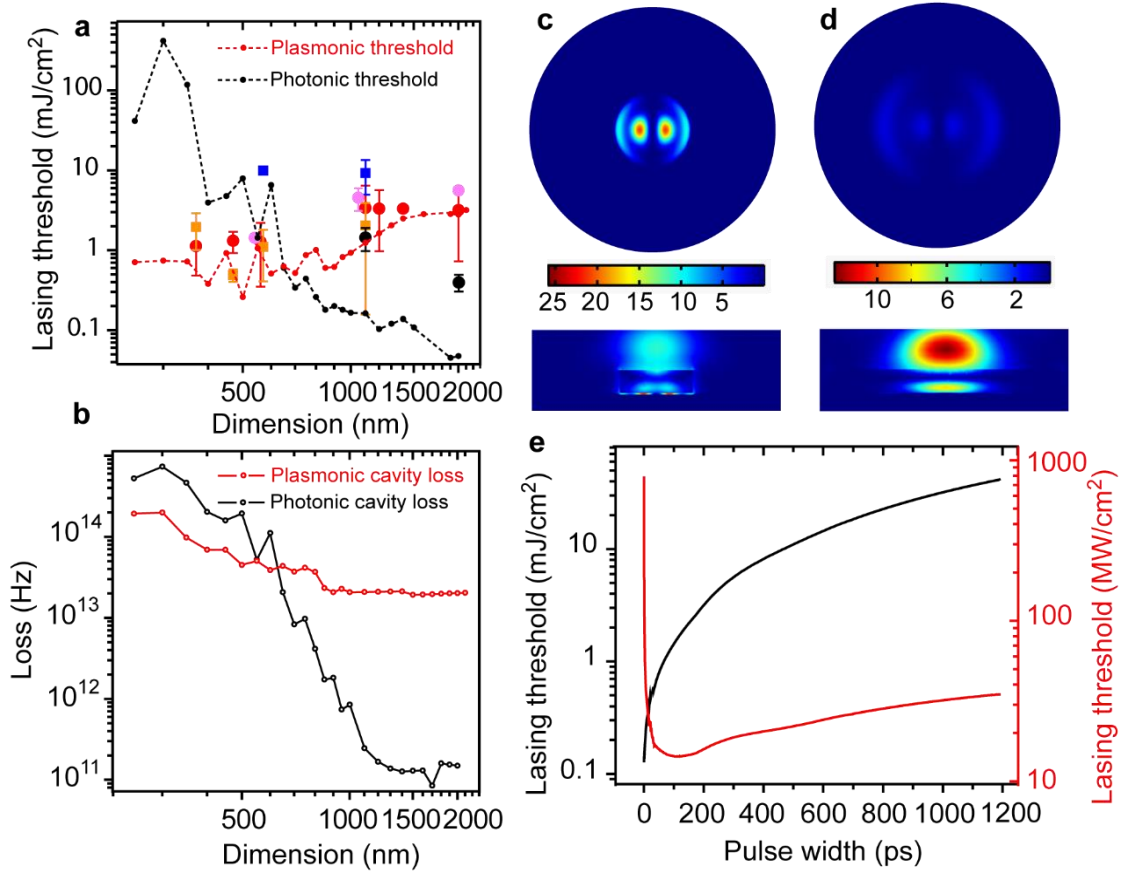

**Figure 5. Lasing threshold dependence on enhanced absorption, enhanced stimulated emission, and pulse duration. (a).** Summary of lasing thresholds obtained on 3D disks and squares in hybrid plasmonic and photonic cases. ● and ● indicate experimental lasing thresholds obtained on hybrid plasmonic disks pumped by 4 ps and 76 ps lasers respectively, while ● corresponds to results acquired from photonic disk cavities pumped by 4 ps. ■ and ■ describe experimental lasing thresholds obtained from hybrid plasmonic squares and photonic cases when pumped by 4 ps, respectively. The lasing thresholds were obtained from three sets of samples and the standard deviation of the thresholds is plotted as the error bar accompanying each data point. -●- and -●- are simulated lasing threshold using rate equations with all parameters from COMSOL simulations for hybrid plasmonic and photonic disk cases. **(b).** –○–, and –○– are COMSOL simulated cavity losses for circular disks of hybrid plasmonic and photonic cases, respectively. The mode of least loss for a given diameter is chosen in the wavelength range of 610 nm to 685 nm, which is why the plots show zigzagged features. **(c-d).** COMSOL simulation of total electric field intensity $|E|^2$ distribution within the disk cavities of 350 nm and 1.1 μm in diameter, respectively, at pump wavelength 570 nm, focused to the sample by an objective of NA=0.9. Top panels show the distributions at the middle of $Al_2O_3$ layer and bottom panels correspond to the distributions along x-z plane. **(e).** Simulated lasing thresholds for a hybrid plasmonic disk cavity as a function of pump pulse width. The cavity loss is chosen to be $5 \times 10^{13}$ Hz, Purcell factor 6, $\beta$ = 0.05 and with non-radiative lifetime 200 ps. Black curve corresponds to lasing threshold in term of pulse energy (left axis) and red curve describes the lasing threshold in term of peak pump intensity (right axis).